\documentclass[aps,prb,preprintnumbers,twocolumn,showpacs]{revtex4}

\usepackage{epsfig}             % Graphical package for EPS files
\usepackage{dcolumn}            % Alignment of decimal point in tables
\usepackage{bm}			% Bold math mode

\newcommand\w{cm$^{-1}$}                % Wavenumber
\newcommand\ai{\textit{ab-initio\/}}    % ab-initio
\newcommand\etal{\textit{et al.}}       % at al.

\begin{document}

%%%%%%%%%%%%%%%%%%%%%%%%%%%%%%%%%%%%%%%%%%%%%%%%%%%%%%%%%%%%%%%%%%%%%%%%%%%%%

%%tth:\def\protect{ }           % \protect
%%%%%%%%%%%%%%%%%%%%%%%%%%%%%%%%%%%%%%%%%%%%%%%%%%%%%%%%%%%%%%%%%%%%%%%%%%%%%
\title{Phonons and structures of tetracene polymorphs at low
temperature and high pressure}

\author{Elisabetta Venuti}
\author{Raffaele Guido Della Valle}
\author{Luca Farina}
\author{Aldo Brillante}
\affiliation{Dipartimento di Chimica Fisica e Inorganica and INSTM-UdR Bologna,
Universit\`a di Bologna, Viale Risorgimento 4, I-40136 Bologna, Italy}
\author{Matteo Masino}
\author{Alberto Girlando}
\affiliation{Dipartimento di Chimica G.I.A.F. and INSTM-UdR Parma,
Universit\`a di Parma, Parco Area delle Scienze, I-43100, Parma, Italy}

\date{\today}

%%%%%%%%%%%%%%%%%%%%%%%%%%%%%%%%%%%%%%%%%%%%%%%%%%%%%%%%%%%%%%%%%%%%%%%%%%%%%
\begin{abstract}
Crystals of tetracene have been studied by means of lattice phonon Raman
spectroscopy as a function of temperature and pressure. Two different phases
(polymorphs I and II) have been obtained, depending on sample preparation and
history. Polymorph I is the most frequently grown phase, stable at ambient
conditions. Application of pressure above 1 GPa yields polymorph II, which is
also obtained by cooling the sample below 140 K. However, the conditions for
inducing the phase transitions depend on sample preparation and hystory, and
polymorph II can also be maintained at ambient conditions.  We have
calculated the crystallographic structures and phonon frequencies as a
function of temperature, starting from the configurations of the energy
minima found by exploring the potential energy surface of crystalline
tetracene. The spectra calculated for the first and second deepest minima
match satisfactorily those measured for polymorphs I and II,
respectively. The temperature dependence of the spectra is described
correctly. All published x-ray structures, once assigned to the appropriate
polymorph, are also reproduced.
\end{abstract}

\pacs{63.20.-e, % Phonons in crystal lattices
      81.30.Hd, % Solid-solid phase transformations, polymorphism
      78.30.-j} % Infrared and Raman spectra

\maketitle

%%%%%%%%%%%%%%%%%%%%%%%%%%%%%%%%%%%%%%%%%%%%%%%%%%%%%%%%%%%%%%%%%%%%%%%%%%%%%
\section{Introduction} \label{s:intro}

Among molecular organic semiconductors, oligoacenes crystals represent a
subject of increasing experimental and theoretical interest, because their
high carrier transport properties make them likely candidates for
applications in electronic and opto-electronic
devices. \cite{Gundlach02,Butko03,Boer03} New techniques have been exploited
for the growth of acene ultrapure single crystals
\cite{Butko03,Boer03,Laudise98} or thin solid films, \cite{Voigt03} with the
aim of obtaining high quality, well ordered crystalline samples. In fact, the
absence of structural defects is crucial for the achievement of optimal
performances in charge carrier mobilities. \cite{Boer03}

Much effort has been devoted to clarify the polymorphism of
pentacene. \cite{Venuti02,Brillante02,Valle03,Mattheus03,Siegrist01} Starting
from all the published x-ray structures for crystalline pentacene,
\cite{Siegrist01,Mattheus01,Campbell62,Holmes99} we computed the structures
of minimum potential energy, and obtained two local minima of the potential
energy, i.e., two different ``inherent structures'' of mechanical
equilibrium. \cite{Venuti02} This behavior indicated that there were at least
two different single crystal polymorphs of pentacene. The calculations
predicted significant differences between the corresponding Raman spectra of
the lattice phonons, which we checked experimentally, confirming the
existence of two polymorphs. \cite{Brillante02} The correct identity of the
samples, initially assigned only by matching experimental and calculated
spectra, was thus verified directly with x-ray diffraction
measurements. Finally, we obtained theoretical information on the global
stability of the minima by systematically sampling the potential surface of
crystalline pentacene. \cite{Valle03} We found that the two polymorphs
correspond to the two deepest minima. Further deep minima with layered
structures, which might correspond to the thin film polymorphs found to grow
on substrates, were also predicted.

The existence of high temperature (HT), \cite{Campbell62,Holmes99} low
temperature (LT) and high pressure (HP) polymorphs
\cite{Prikhotko66,Vaubel70,Turlet73,Kolendritskii79,Jankowiak79,Sondermann85,Kalinowski78,Rang01}
of tetracene has been known and studied in the past, and has been reported
recently in studies on electronic transport in tetracene single
crystals. \cite{Boer03,Boer04} Phase transitions for this system seem indeed
to occur under variable conditions, depending on sample preparation, history
and cooling speed. \cite{Boer03,Jankowiak79,Sondermann85,Boer04} As the
transformations are sluggish, the sample can show a large temperature range
in which more than one structure is present, and important hysteresis effects
can be observed. Generally, this results in lowered carrier mobilities and
shattering of the crystal upon cooling. Altogether, however, not much is
known about the characteristics of the transitions and the nature of the
polymorphs involved. For instance, it is not clear yet how many phases are
actually formed at low temperature, or whether a LT phase does correspond to
the HP one.

In this paper we address the issue of polymorphism in tetracene with the
methods already successfully used for pentacene polymorphs. First, we provide
lattice phonon Raman spectra obtained by means of a microprobe technique for
two different phases of solid tetracene as a function of both temperature and
pressure. Interfacing optical microscopy to Raman spectroscopy allows for a
detailed mapping of the physical features of each crystalline sample and
probe the conditions under which more phases can be simultaneously
present. Secondly, experimental data are compared with the results of quasi
harmonic lattice dynamics \cite{Ludwig67,Valle96,Valle98} (QHLD) calculations
performed by using either the available crystallographic data,
\cite{Campbell62,Holmes99,Sondermann85} or the theoretical predictions of the
most stable crystal structures for this system, which were obtained by
performing a systematic sampling of the potential energy surface,
\cite{Valle04} as already done for pentacene.

\section{Experimental} \label{s:experimental}

High temperature crystalline tetracene (HT structure) can be obtained by
sublimation under vacuum of the commercial product (Aldrich) as thin
platelets of the length of a few mm and thickness of tenths of
$\mu$m. Samples of microcrystalline powder obtained by a fast sublimation
process may instead contain domains of the low temperature (LT) structure as
physical impurity. Sublimation in an inert atmosphere at reduced pressures
(10--20 kPa of nitrogen or argon) at 493~K, with a procedure similar to that
used to obtain polymorph II of pentacene, \cite{Brillante02} yielded directly
larger amount of physically pure LT phase.

Raman scattering has been detected by using several laser lines, eventually
selecting low energy excitation from a krypton laser tuned at 752.5 nm to
minimize sample fluorescence, due perhaps to some residual chemical
impurities. \cite{Boer03} Raman spectra above 1 GPa were anyway overlapped by
strong fluorescence in all conditions. The spectra have been collected and
analysed by the Jobin Yvon T64000 spectrometer equipped with a liquid
nitrogen cooled CCD detector. Low temperatures $T$ down to 80~K were achieved
in a conventional cryostat (Linkam HFS 91) with a temperature gradient of
10~K/min. High pressures $p$ up to 6 GPa were obtained in a LOTO diamond
anvil cell, \cite{Loto} using perfluorocarbon as pressure medium. Pressures
were measured with the ruby luminescence method. \cite{Piermarini75}

Low $T$ and high $p$ cells were placed on the stage of the microscope
(Olympus BX40) directly interfaced to the spectrometer. The use of 20x and
50x magnification objectives allowed for a spatial resolution of 2.2 and 1.1
$\mu$m, respectively, yielding the possibility to spatially check the
physical purity of crystal polymorphs by mapping the lattice phonon profiles
along the sample surface. \cite{Brillante02}

\section{Computational methods}\label{s:calculations}

The crystal structures and vibrational frequencies of tetracene have been
calculated with the same procedure used for pentacene,
\cite{Brillante02,Valle03,Masino02} following a well assessed
treatment. \cite{Valle01,Girlando00} We first compute {\ai} molecular
geometry, atomic charges, vibrational frequencies and cartesian eigenvectors
of the normal modes for the isolated tetracene molecule. This is done with
the Gaussian98 program \cite{Frisch98} (Rev.~A.5), using the 6-31G(d) basis
set combined with the B3LYP exchange correlation
functional. \cite{Frisch98,Lee88} The vibrational frequencies are scaled by
the factor of 0.9613 recommended \cite{Rauhut95,Scott96} for the combination
of B3LYP and 6-31G(d).

The crystal total potential energy $\Phi$ is given in terms of an atom-atom
Buckingham model, \cite{Califano81} with Williams parameter set IV,
\cite{Williams67} combined with an electrostatic contribution represented by
a set of {\ai} atomic charges. We have chosen the potential derived charges,
\cite{Frisch98} which describe directly the electrostatic potential.

The effects of temperature and pressure are accounted for by computing the
structures of minimum Gibbs energy $G(p,T)$ with a QHLD
method.\cite{Ludwig67,Valle96,Valle98} In this method, where the vibrational
Gibbs energy of the phonons is estimated in the harmonic approximation, the
Gibbs energy of the system is $G(p,T) = \Phi + pV+ \sum_i h \nu_i / 2 + k_B T
~ \sum_i \ln\left[1 - \exp\left(- h \nu_i / k_B T \right) \right]$. Here $V$
is the molar volume, $\sum_i h \nu_i/2$ is the zero-point energy, and the
last term is the entropic contribution. The sums are extended to all phonon
frequencies $\nu_i$. Like pentacene, also tetracene exhibits {\ai} vibrational
frequencies in the energy range of the lattice modes, and the coupling
between lattice and intramolecular vibrations cannot be
neglected. \cite{Valle01,Filippini84} To account for it, we adopt an
exciton-like model, \cite{Girlando00,Califano81} where the interaction
between different molecular coordinates is mediated by the intermolecular
potential which depends directly on the atomic displacements. Since these
correspond to the cartesian eigenvectors of the normal modes of the isolated
molecule, we use the {\ai} eigenvectors and the scaled {\ai}
frequencies. Intramolecular modes above 300 {\w} are not taken into account,
as the coupling is expected to be important only for low frequency
modes. \cite{Masino02}

\section{Temperature dependence of Raman spectra at constant pressure}\label{s:LTS}

For tetracene two distinct bulk crystal structures have been detected by
x-ray diffraction experiments. At room $p$,$T$ conditions tetracene crystal
was found to be triclinic, \cite{Campbell62} space group $P\overline{1}$
($C_i^1$). The unit cell contains two independent molecules located on the
(0,0,0) and ($\frac{1}{2}$,$\frac{1}{2}$,0) inversion sites. Thus, the factor
group analysis for the lattice phonons at $k = 0$ predicts six Raman active
modes of $A_g$ symmetry and three IR active modes of $A_u$ symmetry. More
recently Holmes {\etal} \cite{Holmes99} reported full data for crystalline
tetracene at 183~K. As seen in section \ref{s:calculations}, and unlike what
found for pentacene, \cite{Venuti02,Brillante02} the two experimental
structures of refs. \onlinecite{Campbell62,Holmes99} belong to a single
polymorph. Finally, in 1985 Sondermann {\etal} \cite{Sondermann85} identified
by x-ray diffraction a second triclinic polymorph, for which unit cell
parameters at 140~K were given, but it was not possible to determine whether
the space group was $P\overline{1}$ or $P{1}$.

Room $T$ Raman spectra of crystalline tetracene were reported in the late
seventies by Jankowiak {\etal}, \cite{Jankowiak79} who also reported Raman
spectra as a function of $T$. Discontinuous changes in the temperature
dependence of Raman lattice modes and the appearance of new phonon lines at
182 and 144~K were interpreted with the occurrence of two phase transitions,
\cite{Jankowiak79} although some doubts could be harbored about the evidence
of the high temperature (182~K) one. Also, a number of spectroscopic methods
were used to test the occurrence of low temperature polymorphs in tetracene,
\cite{Prikhotko66,Vaubel70,Turlet73,Kolendritskii79,Kalinowski81} yielding a
quite large range of transition temperatures and conditions. The shattering
of the sample upon cooling and large hysteresis effects upon heating the low
$T$ structure have been observed as a consequence of the phase
transition. \cite{Jankowiak79,Sondermann85}

\begin{figure}[b] % Figure 1
\epsfig{file=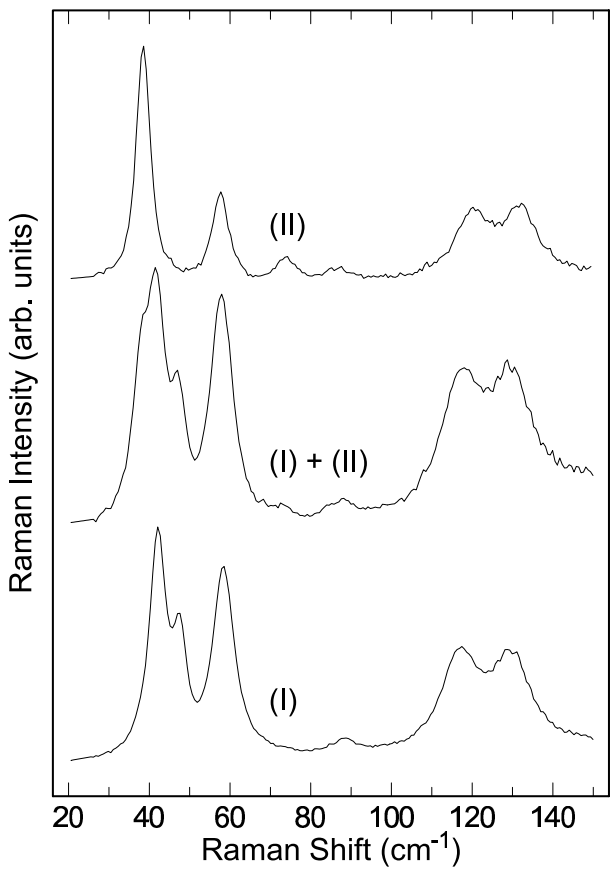,width=8cm}
\caption{\label{f:spectraac}Raman spectra at ambient $p$, $T$ for tetracene
polymorphs. Bottom: Polymorph I; Centre: Mixed phase; Top: polymorph II.}
\end{figure}

\begin{figure}[t] % Figure 2
\epsfig{file=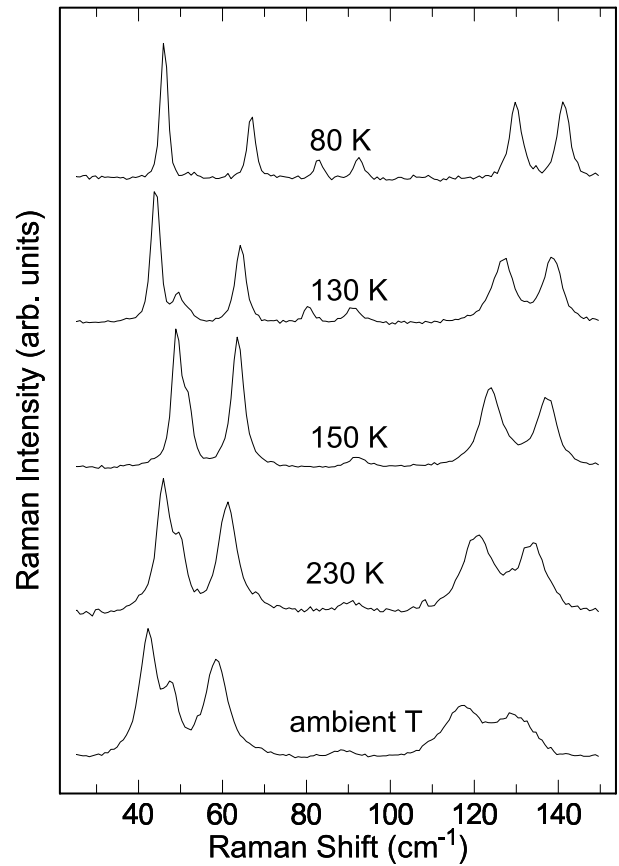,width=8cm}
\caption{\label{f:spectralt}Raman spectra of tetracene crystal as a function
of $T$ at ambient $p$.}
\end{figure}

\begin{table*}[t] % Table I
\caption{Raman wavenumbers ({\w}) of the lattice and intramolecular modes up
to 240~{\w} for polymorphs I and II of tetracene. We report the experimental
$A_g$ wavenumbers, the corresponding minimum $G(p,T)$ calculations, and, for
the intramolecular modes, the {\ai} frequency and symmetry of the parent mode
in the isolated $D_{2h}$ molecule.}
\vspace{5pt}
\hrule
\begin{minipage}{12cm}
\begin{ruledtabular}
\begin{tabular}{cccccccccccccccccccccc}
 \multicolumn{2}{c}{Polymorph {I}}
                         && \multicolumn{5}{c}{Polymorph {II}} \\
 \multicolumn{2}{c}{298~K} &~&  \multicolumn{2}{c}{80~K} &~&
                         \multicolumn{2}{c}{298~K} && \multicolumn{2}{c}{\ai}\\
\cline{1-2}  \cline{4-5} \cline{7-8} \cline{10-11}
expt. & calc.  && expt. & calc. && expt. &calc. && freq.  &sym. \\
\hline
42.3  & 36.5   && 46.1  & 31.4  && 38.3  & 24.4                 \\
47.8  & 44.8   && 66.9  & 66.1  && 57.4  & 55.0                 \\
58.5  & 62.0   && 83.2  & 75.6  && 73.0  & 65.4                 \\
88.4  & 88.8   && 93.2  & 94.3  && 86.1  & 81.7                 \\
117.1 & 131.3  && 129.9 & 151.7 && 118.8 & 136.1                \\
129.8 & 139.1  && 141.2 & 158.6 && 130.4 & 144.3                \\*[2pt]
      & 175.2  && 166.7 & 172.0 && 165.1 & 164.6                \\*[-6pt]
168.2 &        &&       &       &&       &        &$\}$&146.5 & $b_{1g}$   \\*[-6pt]
      & 176.4  && 172.9 & 184.8 && 171.2 & 175.3                        \\
211.2 & 222.1  && 214.9 & 230.3 && 213.3 & 221.3                \\*[-6pt]
      &        &&       &       &&       &        &$\}$&188.2 & $b_{2g}$  \\*[-6pt]
217.0   & 225.3  && 218.6 & 231.7 && 217.8 & 223.0              \\
\end{tabular}
\end{ruledtabular}
\end{minipage}
\label{t:freq1}
\end{table*}

With he aim of rationalizing the situation, we started the experiments by
measuring Raman spectra at room $p$, $T$ in the wavenumber range 20--300~{\w}
for samples grown by sublimation in a variety of ways, as described in
section \ref{s:experimental}. Depending either on the method of preparation
or history of the samples, two different phonon patterns can be observed even
at ambient conditions. To clarify the issue, we report in Figure
\ref{f:spectraac} (bottom trace) the Raman spectrum at 298~K of the thin
tetracene platelets typically obtained by sublimation under vacuum. In the
same Figure (top trace) we also show the spectrum at the same temperature of
samples grown in an inert atmosphere at reduced pressure. Whereas the former
spectrum agrees with that reported \cite{Jankowiak79} by Jankowiak {\etal}
for the HT structure (hereafter called polymorph I), the latter is instead
found to overlap the spectra of samples recovered at ambient conditions after
both LT and HP cycles, as it will be shown in the following. Therefore, these
samples correspond to another phase, which will be called polymorph II. Most
microcrystalline specimens (Figure~\ref{f:spectraac}, centre trace) display
the typical bands of polymorph I with additional features of variable
intensity, which can be attributed to different amount of polymorph II
present as physical impurity.

The phonon wavenumbers for polymorphs I and II at room $T$ are given in Table
\ref{t:freq1}. Since the number of Raman bands observed for polymorph II is
always six, as for polymorph I, it is likely that also polymorph II belongs
to the $P\overline{1}$ space group symmetry. As already revealed by the
spectra of Figure \ref{f:spectraac}, the phonon pattern of the two polymorphs
differs especially in the lowest frequency region. Polymorph I displays three
closely grouped bands in the range 42--58~{\w}, while polymorph II shows
three evenly spaced bands in the range 38--73~{\w}. We remark that this
finding closely resembles what already reported for polymorphs $\textbf{C}$
and $\textbf{H}$ of pentacene, \cite{Brillante02} respectively.

For the low temperature measurements we chose the thin platelets of
physically pure polymorph I, to assure that the starting material for the
temperature cycling was physically homogeneous and belonging to the structure
thermodynamically stable at ambient conditions. Selected spectra recorded on
decreasing temperature in the range 298--80~K are shown in Figure
\ref{f:spectralt}. No discontinuities were seen in our samples down to 140~K.
At 130~K the abrupt appearance of the pattern typical of polymorph II is
observed in the low frequency phonon region. The spectral changes are either
accompanied or preceded by a cracking of the crystal. However, unlike what
reported by Jankowiak {\etal}, \cite{Jankowiak79} there is no hint of an
intermediate crystal modification occurring in the range 180--140~K, even
after repeated temperature cycling on several different specimens. The phase
transformation is clearly completed at 80~K, where no features of the high
temperature spectrum remain. The phonon wavenumbers of the 80~K spectrum of
Figure \ref{f:spectralt} are reported in Table \ref{t:freq1}, and should be
compared with the (incomplete) spectrum at 77~K of
ref. \onlinecite{Jankowiak79}.

\begin{figure}[b!] % Figure 3
\epsfig{file=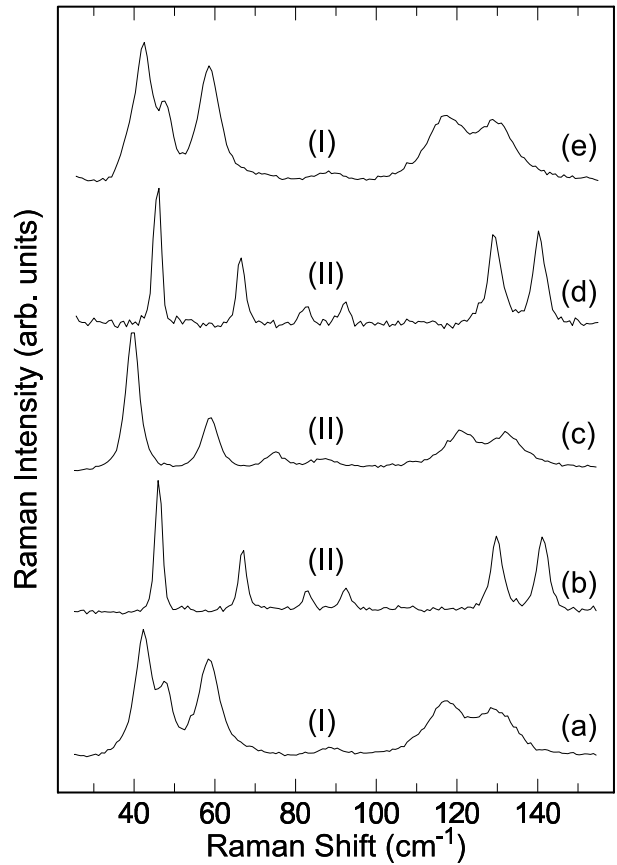,width=8cm}
\caption{\label{f:hysteresis}Raman spectra of a single sample of pentacene
subjected to repeated temperature cycling: (a) starting sample of polymorph I
at 298~K; (b) the sample is transformed to polymorph II by cooling at 80~K;
the features of polymorph II are retained both after (c) returning to ambient
$T$ and (d) cooling to 80 K; (e) polymorph I at 298~K is obtained again after
annealing II at 400~K.}
\end{figure}

Interesting information can be obtained by repeated temperature cycling, as
shown in Figure \ref{f:hysteresis}, where we display a sequence of spectra of
a single sample. After the temperature transition has occurred (Figure
\ref{f:hysteresis} a,b), a large hysteresis is documented by the persistence
of polymorph II upon heating up to 298~K, where all spectral features of this
phase are still retained on the time scale of the experiment, as shown in
Figure \ref{f:hysteresis} c. Note that the latter spectrum overlaps the top
one of Figure~\ref{f:spectraac}. This hysteresis effect allowed us to measure
the temperature dependence of the phonon bands for polymorph II by cooling
the sample over the same $T$ range of polymorph I (Figure \ref{f:hysteresis}
d). Only by heating again polymorph II up to 320~K the conversion to
polymorph I begins, although the process could be completed only by annealing
at 400~K (Figure \ref{f:hysteresis} e). The data will be compared with
calculations in Section \ref{s:cres}.

\begin{figure}[b] % Figure 4
\epsfig{file=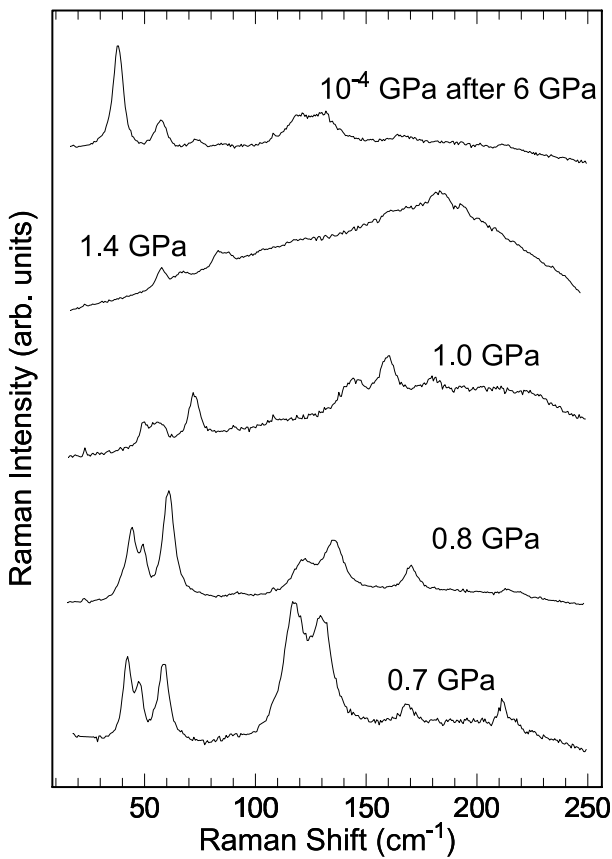,width=8cm}
\caption{\label{f:hpspectra}Raman spectra of crystal tetracene as a function
of $p$ at ambient $T$, showing the transition from polymorph I to polymorph
II in the sample recovered after compression.}
\end{figure}

To summarize, two polymorphs have been clearly identified by the Raman
analysis as function of temperature. Polymorph I is the most frequently
grown, and it is the form stable at room $T$. Polymorph II is the form
obtained by lowering temperature below 140~K. However, it can be obtained as
a (metastable) phase also at room $T$, either by sublimation at 493~K at
reduced pressure in an inert atmosphere or by bringing back to room $T$
samples cooled down to 80~K.

\section{Raman spectra under pressure}\label{s:HPS}

High pressure induced transformations in tetracene crystals were observed as
discontinuous changes of the Davydov splitting in the electronic absorption
spectrum, \cite{Kalinowski78} spatial anisotropy of the magnetic field effect
on fluorescence \cite{Kalinowski76} and recently by studying the
photoconductivity of tetracene single crystals. \cite{Rang01} Mechanical
stress induced by sample grinding was reported to produce a mixture of
different phases. \cite{Jankowiak79,Sondermann85} So far, no Raman spectra
were reported for the pressure induced transformation, and no satisfactory
characterization was performed for the HP phase. Starting from polymorph I,
we have tried to record Raman spectra as a function of $p$ over the range
0--6 GPa. However, in all samples and for all measuring conditions a strong
fluorescence emission hides the Raman scattering above 1.4 GPa, and even in
lower pressure regimes the detection of all bands of the spectrum turns out
to be quite difficult, as shown in Figure~\ref{f:hpspectra}. 
Raman spectra of our samples show that even at 1.0 GPa there
is no sign of phase change. Therefore, the onset of the
pressure-induced phase transition is well above the value (around 0.3~GPa)
previously reported \cite{Kalinowski78,Rang01,Kalinowski76}  with
different experimental techniques.
In any case, the investigation of the samples recovered at ambient $p$
clearly shows that a complete transformation has taken place, as the spectrum
displays all the bands polymorph II and none of polymorph I (Figure
\ref{f:hpspectra}). Therefore, polymorph II can be obtained from polymorph I
also by pressure cycling, in addition to temperature cycling described in the
previous Section.

\begin{figure}[t] % Figure 5
\epsfig{file=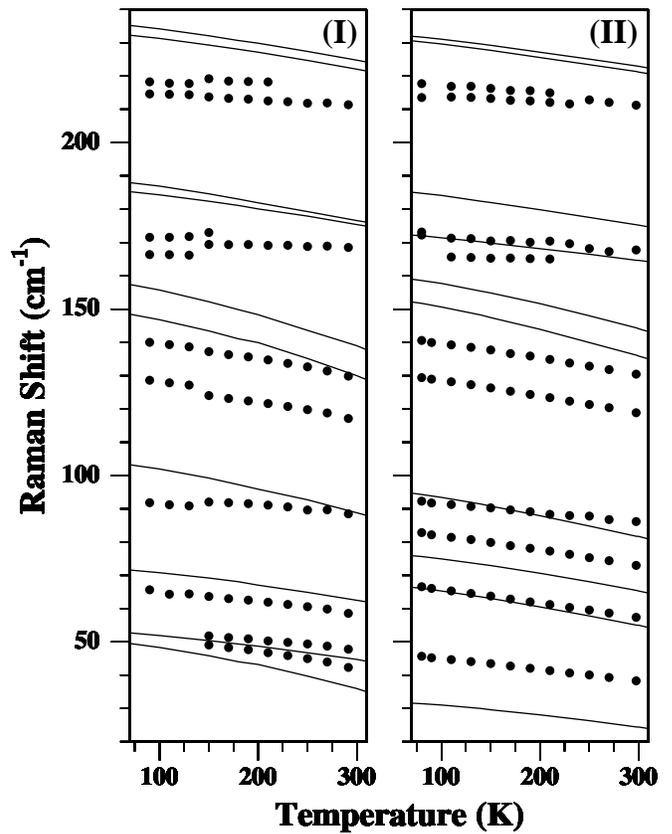,width=8.6cm}
\caption{\label{f:wavet}Phonon wavenumbers {\it vs} temperature for
polymorphs I (left) and II (right) of tetracene. Circles: Raman
experiments. Lines: calculations for $A_g$ phonons.}
\end{figure}

\begin{table*}[t] % Table II
\caption{Lattice parameters of tetracene. The experimental structures of
refs. \protect\onlinecite{Campbell62,Holmes99,Sondermann85} are compared to
the minimum $\Phi$ structures and to the minimum $G(T)$ structures calculated
at the same temperature $T$ (K) of the experiments. Energies are in
kcal/mole, unit cell axes $a$, $b$, $c$ are in {\AA}, angles $\alpha$,
$\beta$, $\gamma$ in degrees, and cell volumes $V$ in {\AA}$^3$.}
\begin{ruledtabular}
\begin{tabular}{llcddddddllllr}
Structure                  & $T$ &  Energy    & a      & b    & c &\alpha & \beta & \gamma & V \\
\hline
\\*[-6pt]
\textbf{Polymorph I} \\
Expt.  $ \protect\cite{Holmes99}$
                           & 180 &            & 6.0565 & 7.8376 & 12.5523 & 101.275 & 99.453  & 94.208 & 572.968 \\ %tholm.ion
Expt.  $ \protect\cite{Sondermann85}$
                           & 293 &            & 6.06   & 7.91   & 12.62   & 101.87  & 99.23   & 94.09  & 581.64  \\ %Sondermann85
Expt.  $ \protect\cite{Campbell62}$
                           & 298 &            & 6.03   & 7.90   & 12.70   & 101.68  & 98.65   & 93.70  & 582.85  \\ %tcamp.ion
Calc.  ${\rm min.}~\Phi$\footnote{Minimum $\Phi$
structure for polymorph I computed starting from the structure of either
ref. \protect\onlinecite{Campbell62} or
ref. \protect\onlinecite{Holmes99}.}
                           &     & $-36.2612$ & 5.8136 & 7.7098 & 12.5972 & 101.310 & 98.272  & 93.548 & 545.537 \\ %tholmC0--99-0.00
Calc.  ${\rm min.}~\Phi$\footnote{Deepest
potential energy minimum. \protect\cite{Valle04}}
                           &     & $-36.2613$ & 5.8133 & 7.7085 & 12.6008 & 101.335 & 98.266  & 93.542 & 545.538 \\ %205006--99-0.00
Calc.  ${\rm min.}~G$      & 180 & $-37.9681$ & 5.8415 & 7.8507 & 12.6665 & 101.353 & 98.437  & 93.569 & 560.855 \\ %tholmC7-180-0.00
Calc.  ${\rm min.}~G$      & 298 & $-42.4059$ & 5.8745 & 7.9384 & 12.7224 & 101.281 & 98.471  & 93.568 & 572.940 \\ %tholmC7-298-0.00
\\*[-6pt]
\textbf{Polymorph II} \\
Expt.  $ \protect\cite{Sondermann85}$
                           & 140 &            & 5.99   & 7.74   & 12.32   & 101.30  & 100.74  & 94.0   & 546.78  \\ %sondermann85
Calc.  ${\rm min.}~\Phi$\footnote{Second deepest
potential energy minimum. \protect\cite{Valle04}}
                           &     & $-35.9089$ & 5.9411 & 7.5882 & 12.8064 & 106.123 & 97.980  & 85.594 & 548.818 \\ %205008--99-0.00
Calc.  ${\rm min.}~G$      & 140 & $-36.5954$ & 5.9646 & 7.6820 & 12.8834 & 106.347 & 98.017  & 85.638 & 560.497 \\ %205008-140-0.00
\end{tabular}
\end{ruledtabular}
\label{t:structures}
\end{table*}

\section{Computational results}\label{s:cres}

We now describe the computational results, which help to definitely clarify
the nature of the two tetracene polymorphs identified through Raman
spectroscopy. In an early stage of our work, we considered separately the two
complete experimental structures. \cite{Campbell62,Holmes99} Each structure
was modeled starting from its experimental molecular arrangements, by
replacing the experimental molecular geometries with the {\ai} one. We thus
discovered that these two structures, measured at different $T$, actually map
into the same potential minimum, having identical energies and unit cells
(Table \ref{t:structures}). Therefore they correspond to a single phase,
stable at room $T$, which we identify with polymorph I. Accordingly, the
subsequent calculations for the $T$ dependence of structure and dynamics for
this polymorph were all performed starting from the data of Holmes {\etal}
\cite{Holmes99} and {\ai} molecular geometry. In Table \ref{t:structures}, we
compare the lattice parameters of the experimental structures of
refs. \onlinecite{Campbell62,Holmes99,Sondermann85} to the calculated
parameters of the structures at the minimum of $\Phi$ and at the minima of
$G(p,T)$. The latter were calculated at ambient pressure, and at the same
temperatures of the experiments. As the various structures of the literature
were not reported in the same standard crystallographic frame, we directly
compare the equivalent reduced cells. \cite{Santoro70}

In Table \ref{t:structures} we also report the minimum $\Phi$ lattice
parameters of the structures which are theoretically predicted to correspond
to the two deepest minima on the potential energy surface. These have been
identified by performing a systematic search for the minima of the potential
energy hypersurface. Following the methods used for pentacene
\cite{Valle03}, we used a Quasi Monte Carlo sampling scheme to generate
several thousands of different initial structures. Starting from each
structure, we then minimized the total potential energy by adjusting the
cell axes angles, positions and orientations of the molecules. The technical
details of the calculations, together with information on the overall
distribution of minima, will be reported in a separate paper. \cite{Valle04}
Here we observe that the two deepest minima present triclinic lattice, space
group $P\overline{1}$ ($C_i^1$), and differ mainly for the orientation of
the two molecules in the unit cell. As shown in Table \ref{t:structures},
polymorph I of refs.  \onlinecite{Campbell62,Holmes99,Sondermann85} maps
very accurately onto the deepest minimum of the potential energy
surface. The table also shows that the minimum $\Phi$ structure for
polymorph I reproduces reasonably well the experimental lattice parameters,
with residual differences $\approx$ 3\% for the unit cell axes and
angles. The computed cell volume is $\approx$~5\% smaller than the
experimental one at room temperatures. This discrepancy, which decreases
upon cooling, is partly due to the thermal expansion, totally neglected in
the minimum $\Phi$ calculations. In fact, as it can be seen from the results
of the minimum $G(p,T)$ calculations, including vibrational effects brings
the calculated volumes within 2\% of the experimental ones, or better, and
reproduces correctly the thermal expansion of phase I. The experimental
volume expands by 1.7\% from 180 to 298~K, to be compared with a calculated
expansion of 2.1\%. Finally, very good is the agreement for the experimental
sublimation heat (unspecified phase) $\Delta_{\rm sub} H = 34.4 \pm 1.2$
kcal/mol, \cite{Dekruif80} which is to be compared with the Gibbs energy
calculated at 0~K, $G(0) \approx 35.1$ kcal/mol.

A totally reliable comparison between our computed structures and the
measurements for polymorph II is not currently feasible, since only the
experimental cell parameters, and no atomic coordinates, are given in
ref. \onlinecite{Sondermann85}. An accurate comparison is also impossible
for the theoretical structures predicted in the same paper on lattice energy
considerations. Note, however, that the prediction of a crystal symmetry
lowering from $P\overline{1}$ to $P1$ is not supported by our Raman
measurements.

In our calculations, we have found that the minimum
$G(p,T)$ structure computed at 140~K starting from the second deepest
minimum of the potential energy surface \cite{Valle04} favorably compares
with the experimental cell parameters \cite{Sondermann85} of polymorph II
at the same $T$. The comparison is reported at the bottom of Table
\ref{t:structures}, and although the agreement with the experiment is with
no doubt worse than for polymorph I, the association appears justified,
and is well supported by the calculated phonon spectrum discussed below.

The experimental Raman frequencies recorded as a function of $T$ for
polymorphs I and II are compared to the corresponding minimum $G(p,T)$
calculations in Figure \ref{f:wavet}. The data for polymorph I are compared
to the results obtained starting from the structure by Holmes \cite{Holmes99}
{\etal}. Instead, the data for polymorph II are compared to those obtained by
starting from the second theoretical deepest minimum. \cite{Valle04} Lattice
and intramolecular modes are both shown. In Table \ref{t:freq1} we report the
calculated Raman wavenumbers for the two structures at 298~K, along with
their experimental values. For polymorph II we also give the values at 80~K.

For both polymorphs the first six Raman modes are almost fully intermolecular
in character, with negligible intramolecular contributions. In fact, the
lowest {\ai} vibration of $g$ symmetry in tetracene is calculated at
146.5~{\w}, and is weakly coupled only with the highest frequency lattice
mode, which displays a intramolecular contribution of about 10\% at
80~K. Therefore, in the Raman spectra tetracene behaves as an apparent
rigid-body while the coupling is expected to be important for ir active modes
of $u$ symmetry above 100~{\w}, as already remarked by Filippini and
Gramaccioli. \cite{Filippini84}

The temperature dependence of the phonon frequencies calculated for phases I
and II (Figure \ref{f:wavet}), also agrees well with the corresponding
experimental results. As expected, both experiments and calculations shows
that varying the temperature affects the low frequency lattice modes much
more than the purely intramolecular modes above 150~{\w}.
The differences between the experimental and computed temperature dependence
of the frequencies, especially noticeable for the high frequency modes, are
attributed to the anharmonic frequency shifts,\cite{Valle83} neglected in
these calculations, and to defects in the potential model.

To summarize the results of the phonon dynamics analysis, we point out again
(cf. section \ref{s:LTS}) that the patterns of the experimental frequencies
for the lattice modes of polymorphs I and II at room $T$ are clearly
distinguishable, and are well matched by the patterns calculated for the
experimental structure \cite{Holmes99} and for the second theoretical
deepest minimum, respectively. In particular, the computations correctly
predict the three closely spaced modes around 48~{\w} for polymorph I, and
reproduce the more widely spread bands between 38 and 73~{\w} for polymorph
II. Therefore we can associate polymorph I and II with the two deepest
minima found in the potential energy surface.\cite{Valle04} Our calculations
actually predict that polymorph I is the more stable (at 0 K) and denser
phase, whereas experimentally the opposite is true.  Moreover, we do not
find theoretical evidence of the phase transition (crossing of $G$ values)
as function of either $T$ or $p$. The same kind of problem has been found in
calculations for pentacene.\cite{Masino02,Verlaak03} We could fine tune the
atom-atom potential to yield the correct phase ordering. On the other hand,
the calculated energy difference between the two phases is very small, less
than 0.5 kcal/mole, which is the typical accuracy of this kind of
calculations. We therefore think that attempts to improve the potential
would be unjustified at the present stage.

\section{Discussion and conclusions}\label{s:conclusions}

In this paper we have explored the $p$,$T$ phase diagram of crystalline
tetracene.  By combining Raman spectroscopy with computational methods, we
have clarified several issues related to the crystalline phases of
tetracene. Tetracene crystallizes into two different polymorphs, polymorph I
and II, the former being the most frequently grown phase, stable at ambient
conditions. Whereas the crystalline structure of polymorph I is well known,
\cite{Campbell62,Holmes99} we suggest a likely structure for polymorph II,
stable at low $T$ and high $p$. The structures of polymorph I and II are very
similar to the structures of polymorph {\bf C} and {\bf H} of pentacene,
respectively.\cite{Masino02} All structures are triclinic, space group
$P\overline{1}$ ($C_i^1$), with two molecules per unit cell residing on
symmetry unrelated inversion centers. In polymorph I and {\bf C}, the long
molecular axis is roughly pointing along the [0,0,1] direction ($c$ axis), in
polymorphs II and {\bf H}, which are the denser phases, the molecules are
more inclined, and point towards the [1,1,$-1$] direction.  Raman
spectroscopy in the lattice phonon region has indicated that the denser
polymorph II of tetracene is stable only at low temperature (below 140 K) or
high pressure (well above 1 GPa), whereas polymorph {\bf H} of pentacene is
the stable phase at ambient pressure.  No additional phases have been
detected, so early reports may have been affected by temperature induced
strains and sample impurities/imperfections.\cite{Kalinowski78}

Finally, we have verified the possibility of obtaining either or both
tetracene polymorphs at ambient $p$,$T$ conditions, depending on
sample preparation. Obtaining pure polymorph II at ambient conditions
can be very important, as we expect that a denser phase should exhibit
larger bandwidths and mobilities. The two tetracene phases are very
similar in energy, and as it happens for pentacene, crystalline
samples may show phase inhomogeneities. The two polymorphs can be
easily identified through Raman spectroscopy in the lattice phonon
region, and possible phase inhomogeneities are detectable this
way. Raman spectroscopy thus represents a convenient and reliable tool
for checking crystal quality, contribuiting to improve the
performances of tetracene-based devices.

%%%%%%%%%%%%%%%%%%%%%%%%%%%%%%%%%%%%%%%%%%%%%%%%%%%%%%%%%%%%%%%%%%%%%%%%%%%%%
\begin{acknowledgments}
Work supported by the Consorzio Interuniversitario Nazionale
per la Scienza e Tecnologia dei Materiali (I.N.S.T.M., PRISMA2002 project)
and by the Italian Ministero Istruzione, Universit\`a e Ricerca
(M.I.U.R., FIRB project).
\end{acknowledgments}

\end{document}